\pdfoutput=1
\documentclass[runningheads]{llncs}
\usepackage{graphicx}
\usepackage{hyperref}
\usepackage{xcolor}
\usepackage{longtable}
\usepackage{array}

\begin{document}
\title{Analyzing Wikidata Transclusion on English Wikipedia}
\author{Isaac Johnson}
\authorrunning{I. Johnson}
\institute{Wikimedia Foundation
\email{isaac@wikimedia.org}}
\maketitle              % typeset the header of the contribution
\begin{abstract}
Wikidata is steadily becoming more central to Wikipedia, not just in maintaining interlanguage links, but in automated population of content within the articles themselves. It is not well understood, however, how widespread this transclusion of Wikidata content is within Wikipedia. This work presents a taxonomy of Wikidata transclusion from the perspective of its potential impact on readers and an associated in-depth analysis of Wikidata transclusion within English Wikipedia. It finds that Wikidata transclusion that impacts the content of Wikipedia articles happens at a much lower rate (5\%) than previous statistics had suggested (61\%). Recommendations are made for how to adjust current tracking mechanisms of Wikidata transclusion to better support metrics and patrollers in their evaluation of Wikidata transclusion.
\end{abstract}
\section{Introduction}
Wikidata is steadily becoming more central to Wikipedia, not just in maintaining interlanguage links, but in automated population of content within the articles themselves. This transclusion of Wikidata content within Wikipedia can help to reduce maintenance of certain facts and links by shifting the burden to maintain up-to-date, referenced material from each individual Wikipedia to a single repository, Wikidata.

Current best estimates suggest that, as of August 2020, 62\% of Wikipedia articles across all languages transclude Wikidata content. This statistic ranges from Arabic Wikipedia (arwiki) and Basque Wikipedia (euwiki), where nearly 100\% of articles transclude Wikidata content in some form, to Japanese Wikipedia (jawiki) at 38\% of articles and many small wikis that lack any Wikidata transclusion. English Wikipedia, the largest language edition, sits in the middle at 61\% articles having some form of Wikidata transclusion.\footnote{\url{https://wmdeanalytics.wmflabs.org/WD_percentUsageDashboard/}}

The challenge in interpreting these numbers, however, is that they are based on a database table called wbc\_entity\_usage\footnote{\url{https://www.mediawiki.org/wiki/Wikibase/Schema/wbc_entity_usage}} that was not built to provide insight into the nature of Wikidata transclusion. Specifically, the table serves to support the Mediawiki software in updating its cache of rendered Wikipedia articles when updates are made to transcluded Wikidata items. In many cases, however, the data is very coarse and gives little indication as to how the Wikidata content appears in the Wikipedia article. Anecdotally, Portuguese Wikipedia (ptwiki) has a featured list article\footnote{\url{https://pt.wikipedia.org/wiki/Lista_de_mortos_e_desaparecidos_pol\%C3\%ADticos_na_ditadura_militar_brasileira}} that is generated purely from Wikidata content using a template\footnote{\url{https://en.wikipedia.org/wiki/Template:Wikidata_list}} and many languages make extensive use of Wikidata-powered infoboxes whereas, by consensus,\footnote{\url{https://en.wikipedia.org/wiki/Wikipedia:Requests_for_comment/Wikidata_Phase_2}} transclusion of Wikidata into infoboxes and article content in English Wikipedia (enwiki) is quite restricted.

The goal of this research is to contextualize the coarse Wikidata transclusion numbers so as to better understand the diversity of ways in which Wikidata is being used within Wikipedia and how these different ways might impact Wikipedia readers. Informed by design decisions and community consensus, a taxonomy of Wikidata transclusion is developed and applied to English Wikipedia via a mixture of qualitative coding and quantitative analyses. Specifically, this research makes the following contributions:
\begin{itemize}
    \item \textbf{Taxonomy of transclusion}: a framework is provided to categorize different types of Wikidata transclusion based upon their perceived potential impact on readers.
    \item \textbf{Analysis of transclusion on English Wikipedia}: it is determined that the vast majority of the 61\% of articles with evidence of Wikidata transclusion on English Wikipedia can be classified as low-importance and only 5\% of articles on English Wikipedia have medium- or high-importance Wikidata transclusion.
\end{itemize}

\section{Background}
Relatively little research has focused on how Wikidata is transcluded within Wikipedia, with most of the focus of studies of Wikidata (outside of the large body of work building tools with Wikidata) either being on Wikidata as an isolated entity---e.g., detecting vandalism on Wikidata~\cite{sarabadani2017building}, estimating item completeness~\cite{balaraman2018recoin}---or studying the distribution of knowledge across languages via labels and Wikipedia sitelinks~\cite{kaffee2017glimpse,konieczny2018gender}. While this past research is not directly related to the question of in what ways Wikidata is transcluded in Wikipedia articles, the results of this analysis do provide insight into one measure of a Wikidata item's importance, a relevant aspect of understanding completeness and prioritizing work such as vandalism detection. It has been noted, as well, that Wikidata transclusion introduces challenges to Wikipedia editors ability to patrol changes to articles~\cite{morgan2019patrolling}.

% mention https://en.wikipedia.org/wiki/Wikipedia:Mbabel/Templates
\subsection{Wikidata Transclusion}
Wikidata transclusion is tracked via the wbc\_entity\_usage database table\footnote{\url{https://www.mediawiki.org/wiki/Wikibase/Schema/wbc_entity_usage}}, the intention of which is to keep the cache of parsed Wikipedia articles up-to-date. It thus tracks which Wikidata items, and which aspects of these items, are transcluded in Wikipedia articles. It is based on what Wikidata content is requested when rendering a Wikipedia article, however, which is a superset of what Wikidata content actually is transcluded (and displayed) in an article.

There are a variety of mechanisms through which Wikidata transclusion happens within Wikipedia articles such as templates that depend on Lua modules or more specific parser calls.\footnote{\url{https://en.wikipedia.org/wiki/Wikipedia:Wikidata\#Inserting_Wikidata_values_into_Wikipedia_articles}} While the end result should not depend on the method employed, the exact method used does determine how the transclusion is recorded in the wbc\_entity\_usage table. In particular, whereas the specific parser calls often allow the table to record the specific statements used within an article, the more general Lua modules often trigger a much broader usage aspect (``some statements'') that obscures which particular properties are being transcluded.

Calculating which articles have some form of Wikidata transclusion through the wbc\_entity\_usage table, then, is a very coarse measure of what Wikidiata content is actually included in a Wikipedia article. Improving the insights that can be gleaned from this table would require substantial reengineering of how the table works. This research, however, does provide some insight into how to interpret the data in this table as well as offer an alternative, richer, though less automatic, framework for evaluating Wikidata transclusion within Wikipedia.

\section{Methods}

\subsection{Taxonomy of Wikidata Transclusion}
Not all forms of Wikidata transclusion have the same potential impact. In this research, impact was defined as how strongly it is believed a reader could be affected if the content that was transcluded was incorrect. There are many alternative definitions---e.g., impact on editors' backlog to maintain Wikipedia---that would undoubtedly lead to different classifications. This definition was chosen as it informs how important it is for patrollers to be aware of the transclusion and track changes to the content~\cite{morgan2019patrolling}.

Instances of Wikidata transclusion was classified as either high, medium, or low importance---i.e. impact---depending on three different criteria:
\begin{itemize}
    \item \textbf{Platform}: does the transclusion show up on all platforms---i.e. desktop, mobile web, and Android/iOS apps---or only in certain form factors? Transclusion that shows up on more platforms has a higher potential for impact because more readers can see it. English Wikipedia (like most other language editions) is largely split between desktop and mobile web readers with only a small percentage of pageviews coming from the apps.\footnote{\url{https://stats.wikimedia.org/\#/en.wikipedia.org/reading/total-page-views/normal|bar|2-year|access~desktop*mobile-app*mobile-web|monthly}}
    \item \textbf{Salience}: when transclusion appears, where on the page does it appear and how obvious is it? Content that appears at the top of articles is more likely to be interacted with by readers versus content that appears at the end of articles~\cite{dimitrov2017makes} and thus is more impactful. Links generated by templates generally have low click-through rates~\cite{mitrevski2020wikihist}.
    \item \textbf{Reader impact if viewed}: finally, assuming the reader sees the content, what is the potential impact? Is it a fact that might affect their perception of the topic~\cite{rothshild2019interplay,singer2017we}, a link that could lead to offensive content or disinformation such as media repositories that might contain offensive images, a link that could lead to incorrect content but points to a well-curated knowledge base such as VIAF (libraries) or GNIS (geography) and would be likely recognized as incorrect if followed and not be offensive, or is it just a hidden tracking category\footnote{\url{https://en.wikipedia.org/wiki/Wikipedia:Categorization\#Wikipedia_administrative_categories}} for the article that assists in maintenance of the wiki but has no relevance to the reader?
\end{itemize}

Table~\ref{tab:templates} shows the main types of templates that were encountered on English Wikipedia and how they were coded according to the above taxonomy.

\begin{table*}
\centering
\begin{tabular}{l|lllllp{3cm}<{\raggedright}}
Use-case & Platform & Salience & Impact & Importance & Usage & Example Templates \\
\hline
Infobox & All & High & High & High & Low &  \href{https://en.wikipedia.org/wiki/Template:Infobox_person/Wikidata}{Ib person Wikidata}; \href{https://en.wikipedia.org/wiki/Template:Infobox_company}{Ib company} \\
External Links & All & Medium & Medium & Medium & Medium & \href{https://en.wikipedia.org/wiki/Template:IMDb_title}{IMDb title}; \mbox{\href{https://en.wikipedia.org/wiki/Template:Commons_category}{Commons category}} \\
Search Descriptions & App & High & High & Low & High & Default without shortdesc \\
Metadata & Desktop & Low & Low & Low & High & \href{https://en.wikipedia.org/wiki/Template:Authority_control}{Authority control};  \href{https://en.wikipedia.org/wiki/Template:Taxonbar}{Taxonbar} \\
Tracking Categories & Desktop & Low & Low & Low & High & \href{https://en.wikipedia.org/wiki/Template:Birth_date}{Birth date};  \href{https://en.wikipedia.org/wiki/Template:Coord}{Coord (when hard-coded)} \\
References & All & Low & Low & Low & Medium & \href{https://en.wikipedia.org/wiki/Template:USGS_gazetteer}{USGS gazetteer}; \mbox{\href{https://en.wikipedia.org/wiki/Template:Cite_Q}{Cite Q}} \\\\
\end{tabular}
\caption{Overview of the primary forms in which Wikidata transclusion happens on English Wikipedia along with how each use-case is coded for its importance to readers. ``Ib'' stands for ``Infobox''. Usage refers to roughly how often each use-case appears in English Wikipedia based on the results of this study.}\label{tab:templates}
\end{table*}

\subsection{Qualitative Coding}
One hundred random articles on English Wikipedia were chosen and evaluated to determine what, if any, templates were being recorded as transcluding Wikidata content in the article and categorized into the transclusion categories (low-, medium-, high-importance) described above.\footnote{The data from the wbc\_entity\_usage table can be easily viewed from within the article in the ``Page information'' tab under ``Wikidata entities used in this page''.} An article was categorized with its highest level of transclusion---e.g., an article with a low-importance transclusion and high-importance transclusion would be coded as high-importance transclusion. Only a single researcher evaluated the data given the time-intensive nature of the coding. While there is in theory only a single correct answer for what templates are triggering Wikidata transclusion, in practice it can be difficult to trace back the calls made by templates to determine whether they might be causing the transclusion listed within the wbc\_entity\_usage table. The articles were evaluated in their current form on 31 July 2020.

\subsection{Quantitative Verification}
The final step was to verify the representativeness of the sample of articles coded and prototype an approach to continuing to track Wikidata transclusion without having to re-code a new sample of articles. Specifically, for the templates that were identified as major sources of Wikidata transclusion, a Python script was developed that would detect usage of these templates in wikitext dumps\footnote{\url{https://github.com/geohci/wikidata-transclusion}} and determine whether the template was actually transcluding content or just generating tracking categories based on a series of heuristics.\footnote{Alternative strategies such as selectively ablating templates from a page in a test environment to see how content and tracking changes would be excessively complex because most templates are written to use the Wikidata item of the article containing them and so cannot be evaluated outside of the context of the article they are in.} This included two common metadata templates (Template:Authority control and Template:Taxonbar), all templates that generate external links from Wikidata\footnote{\url{https://en.wikipedia.org/wiki/Category:External_link_templates_using_Wikidata}} as they follow a very similar pattern, and the templates for inserting latitude/longitude coordinates (Template:Coord) and birth date (Template:Birth date and age) as very common Wikidata templates. Redirects to these templates were also included. Infobox templates were not included in this script as there are approximately 100 with varying but low usage\footnote{\url{https://en.wikipedia.org/wiki/Category:Infobox_templates_using_Wikidata}} and each template has its own specific parameters that control Wikidata transclusion. The 1 August 2020 wikitext dumps were used for this evaluation.\footnote{\url{https://dumps.wikimedia.org/enwiki/20200801/enwiki-20200801-pages-articles.xml.bz2}}

\section{Results}
As a baseline, the wbc\_entity\_usage table indicates as of August 2020 that 61\% of articles on English Wikipedia include some form of Wikidata transclusion.\footnote{\url{https://wmdeanalytics.wmflabs.org/WD_percentUsageDashboard/}} In line with this statistic, from the 100 article sample selected to be evaluated (full data in Table~\ref{tab:articlesample} in the Appendix), 59 had some form of Wikidata transclusion recorded in the wbc\_entity\_usage table.

For those 59 articles with some evidence of Wikidata transclusion, however, only 5 had medium- or high-importance Wikidata transclusion on further inspection. Specifically, 2 articles had Wikidata-powered infobox templates, 3 articles had Wikidata-powered external link templates, 21 articles just had metadata templates, and 33 articles just had Wikidata tracking categories. Overlapping with these categories, 54 of the 100 articles did not override the Wikidata description for use in article previews in Search on the official iOS and Android apps.\footnote{\url{https://www.mediawiki.org/wiki/API:Page_info_in_search_results}} This form of transclusion is not tracked by the wbc\_entity\_usage table though.

The quantitative evaluation supported the validity of the transclusion rates seen in the qualitative coding. From all of English Wikipedia (6,125,693 articles), 3\% of articles had external link templates that transcluded Wikidata (only 47.5\% of the 3,132,833 instances found of external link templates that could transclude Wikidata did while the rest provided hard-coded parameters and thus only served to generate tracking categories), 29\% of articles had metadata templates\footnote{This actually tracks closely to the 21\% statistic from the qualitative evaluation as that number is for articles that only had metadata templates but an additional 3\% of articles were found to have metadata templates and higher-importance templates.}, and 27\% of articles were found to just have tracking categories generated via Wikidata templates. For instance, out of the 1,841,505 times in which the coordinates template (displays latitude-longitude coordinates in an article with a link to a map) appears in English Wikipedia, it only uses Wikidata to provide coordinates 2,730 times (0.1\%) and the rest of the template instances have latitude-longitude coordinates hard-coded.\footnote{This is likely the result of English Wikidata determining that existing information should not be overwritten with Wikidata data: \url{https://en.wikipedia.org/wiki/Wikipedia:Requests_for_comment/Wikidata_Phase_2}}

\section{Discussion}
\subsection{Tracking Wikidata Transclusion}
This analysis indicates that the prior measure of Wikidata transclusion in English Wikipedia (61\%) both greatly underestimates and overestimates Wikidata transclusion. Transclusion that affects the content of articles (high- and medium-importance) was determined to be 5\% of articles, indicating that 61\% is far too high of a measure for these types. On the other hand, usage of the Wikidata description in the mobile apps has not been previously tracked and so this 61\% number does exclude some articles that are indeed using Wikidata descriptions within the context of the Wikipedia apps.

Metrics that track how well-incorporated Wikidata is into Wikipedia articles will need more nuanced approaches that capture these distinctions. Two fixes would make a huge impact. First, if standard statements---e.g., date of birth---were tracked separately from identifiers---e.g., VIAF, GNIS IDs---this would separate out metadata templates. This analysis found that metadata templates account for much of Wikidata transclusion. They speak to Wikidata's value as a linked data repository but are a very different use-case than transclusion of content like date of birth or coordinates. Secondly, transclusion that merely generates tracking categories---e.g., \href{https://en.wikipedia.org/wiki/Category:Date_of_birth_not_in_Wikidata}{Category:Date of birth not in Wikidata}---could be tracked separately from transclusion that inserts content into the page. 

Together, these two changes would greatly improve the ease with which the different importance-levels of Wikidata transclusion could be tracked without in-depth qualitative coding wiki by wiki. This would both support more nuanced metrics of Wikidata transclusion---e.g., for tracking progress---but also support much more nuanced filters for patrollers of changes to Wikidata that might affect articles on their watchlist. Currently, patrollers of vandalism can choose to see or not see Wikidata transclusion,\footnote{\url{https://en.wikipedia.org/wiki/Wikipedia:Wikidata\#Recent_changes}} but do not have good tools for distinguishing from transclusion that might have a sizeable impact on an article and therefore warrant checking and changes to Wikidata items that would have no or only a very minimal impact on an article. As these analyses demonstrated, much of Wikidata transclusion could reasonably be considered ``noise'' from a patrolling perspective, especially because the metadata templates are tracked via wbc\_entity\_usage in such a way that any change to any part of the article's associated Wikidata item shows up in recent changes (even if it has no impact on the article whatsoever).

\subsection{Generalization}
English Wikipedia was analyzed in this work as the most accessible Wikipedia for qualitative coding for the researcher. While there are both many wikis with more and with less evidence of transclusion, English Wikipedia does have some very strong policies with regards to Wikidata transclusion that are not necessarily found in other language editions. Further analyses of Wikidata transclusion in other languages would undoubtedly discover different distributions of usage and new types of templates that have not shown up on English Wikipedia.

A different approach to the taxonomy would also change the results. For instance, this taxonomy does not take into account any measure of the importance of the article that is transcluding the Wikidata content---e.g., number of views the article gets~\cite{warncke2015misalignment}, whether it's a biography of a living person\footnote{\url{https://en.wikipedia.org/wiki/Wikipedia:Biographies_of_living_persons}}---which would likely mediate the expected impact of incorrect content being transcluded. Different taxonomies---e.g., focusing on the reduction in editor backlog that would result from a particular instance of Wikidata transclusion---would also likely lead to very different results.

\newpage
\section{Appendix}
\begin{scriptsize}
\begin{longtable}{l p{3.8cm}|llllp{1.6cm}l}
No. & Page Title & S.~only & T.~only & No shortdesc & Metadata & Content & Import. \endhead
\hline
1 & Don't Go Breaking My Heart (Sonic Dream Collective song) & Y &  & Y &  &  & None \\
2 & Donja Raštelica &  & Y &  &  &  & Low \\
3 & Salió el Sol & Y &  & Y &  &  & None \\
4 & Archimantis sobrina &  &  & Y & Taxonbar &  & Low \\
5 & Fiafia & Y &  & Y &  &  & None \\
6 & Daniel Webster &  &  & Y & Auth. Ctrl. &  & Low \\
7 & 1997 European Athletics U23 Championships – Women's 5000 metres & Y &  & Y &  &  & None \\
8 & Urige Buta &  &  &  & Auth. Ctrl. & Sports links & Medium \\
9 & Wolfgang Martin Stroh &  &  &  & Auth. Ctrl. &  & Low \\
10 & Olu Irame &  & Y &  &  &  & Low \\
11 & Travis McGee &  & Y & Y &  &  & Low \\
12 & Online Bible & Y &  & Y &  &  & None \\
13 & Caledonian Brewery &  &  & Y &  & Ib company & High \\
14 & Networking cables &  &  & Y & Auth. Ctrl. & Commons C. & Medium \\
15 & Arachno Creek &  &  &  & Auth. Ctrl. &  & Low \\
16 & Malinvestment & Y &  & Y &  &  & None \\
17 & Steirastoma poeyi &  &  &  & Taxonbar &  & Low \\
18 & 1927 William \& Mary Indians football team & Y &  &  &  &  & None \\
19 & 1919 Uruguayan parliamentary election & Y &  & Y &  &  & None \\
20 & Abbey of the Holy Ghost &  & Y & Y &  &  & Low \\
21 & Lucky Lynx & Y &  & Y &  &  & None \\
22 & South West Pacific (film) &  & Y & Y &  &  & Low \\
23 & Lawrence Townsend &  &  & Y & Auth. Ctrl. &  & Low \\
24 & Anahuarque &  & Y & Y &  &  & Low \\
25 & Platycephalus fuscus &  &  & Y & Taxonbar &  & Low \\
26 & Cambarus williami &  &  & Y & Taxonbar &  & Low \\
27 & Senoussi (cigarette) & Y &  & Y &  &  & None \\
28 & Nuhiji & Y &  &  &  &  & None \\
29 & Songs of Life (The Gufs album) & Y &  &  &  &  & None \\
30 & Darren McKenzie-Potter &  & Y &  &  &  & Low \\
31 & List of aircraft (Sb) & Y &  &  &  &  & None \\
32 & Res Gestae Divi Augusti &  & Y &  &  &  & Low \\
33 & Nightrider (chess) & Y &  &  &  &  & None \\
34 & Afonso V of Portugal &  &  &  & Auth. Ctrl. &  & Low \\
35 & Panikos Hatziloizou &  & Y &  &  &  & Low \\
36 & Bongo Comics & Y &  & Y &  &  & None \\
37 & Gustavia longepetiolata &  &  & Y & Taxonbar &  & Low \\
38 & 1935 SANFL season & Y &  & Y &  &  & None \\
39 & Cat Orgy &  & Y &  &  &  & Low \\
40 & Ulex &  &  &  & Taxonbar &  & Low \\
41 & Roseate & Y &  &  &  &  & None \\
42 & Tourism Corporation of Khyber Pakhtunkhwa & Y &  & Y &  &  & None \\
43 & Fever 121614 & Y &  &  &  &  & None \\
44 & Inside Detroit &  & Y & Y &  &  & Low \\
45 & KKSE & Y &  &  &  &  & None \\
46 & Philip II, Count of Schaumburg-Lippe &  &  & Y & Auth. Ctrl. &  & Low \\
47 & Lyndon, Ohio &  & Y & Y &  &  & Low \\
48 & Principality of Lippe &  & Y & Y &  &  & Low \\
49 & They Are Billions &  & Y & Y &  &  & Low \\
50 & List of fellows of the Royal Society elected in 1973 & Y &  & Y &  &  & None \\
51 & Epiphysis &  &  & Y & Ib Anatomy &  & Low \\
52 & João Valente Bank &  & Y & Y &  &  & Low \\
53 & Clitophon & Y &  &  &  &  & None \\
54 & Johnstown Center, Wisconsin &  & Y &  &  &  & Low \\
55 & List of mergers in Shimane Prefecture & Y &  & Y &  &  & None \\
56 & Highbury Union F.C. & Y &  & Y &  &  & None \\
57 & Babaha &  & Y &  &  &  & Low \\
58 & Metropolis (Anatolia) &  &  & Y & Auth. Ctrl. &  & Low \\
59 & Mat Latos &  & Y & Y &  &  & Low \\
60 & Woodstock Railway & Y &  & Y &  &  & None \\
61 & Megachile osea &  &  &  & Taxonbar &  & Low \\
62 & Public holidays in Japan & Y &  &  &  &  & None \\
63 & Danish Contemporary Bible 2020 &  & Y & Y &  &  & Low \\
64 & Onion River (Minnesota) &  & Y & Y &  &  & Low \\
65 & Zémidjan & Y &  & Y &  &  & None \\
66 & Mena (album) & Y &  &  &  &  & None \\
67 & Volvarina habanera &  &  & Y & Taxonbar &  & Low \\
68 & Flying High (1931 film) &  & Y &  &  &  & Low \\
69 & Myponie Point &  & Y & Y &  &  & Low \\
70 & Elizabeth Mayer &  &  & Y & Auth. Ctrl. &  & Low \\
71 & Notre Dame Catholic School &  & Y &  &  &  & Low \\
72 & List of Canadian Hot 100 top 10 singles in 2011 & Y &  & Y &  &  & None \\
73 & Arthur Koegel & Y &  & Y &  &  & None \\
74 & Lists of Bulgarian films & Y &  &  &  &  & None \\
75 & 6th Parliament of King William III & Y &  & Y &  &  & None \\
76 & DK King of Swing &  & Y & Y &  &  & Low \\
77 & Benjamin Dwyer &  &  & Y & Auth. Ctrl. &  & Low \\
78 & Patton Glacier &  &  & Y &  & USGS gazetteer & Medium \\
79 & McAfee Peak &  & Y & Y &  &  & Low \\
80 & Sommatino &  &  &  & Auth. Ctrl. & Ib Italian Comune & High \\
81 & Pay Takht-e Varzard &  & Y &  &  &  & Low \\
82 & 89 Albert Embankment &  & Y &  &  &  & Low \\
83 & Acalles carinatus &  &  &  & Taxonbar &  & Low \\
84 & Saudia Aerospace Engineering Industries &  & Y & Y &  &  & Low \\
85 & FromeFM & Y &  & Y &  &  & None \\
86 & Clonakilty Cowboys & Y &  &  &  &  & None \\
87 & Gorilla (James Taylor album) & Y &  &  &  &  & None \\
88 & Union Township, Lawrence County, Ohio &  & Y &  &  &  & Low \\
89 & Radio KAOS & Y &  &  &  &  & None \\
90 & Ralph Emery &  & Y &  &  &  & Low \\
91 & Marine Technology Society &  &  &  & Auth. Ctrl. &  & Low \\
92 & ABMA & Y &  &  &  &  & None \\
93 & Sten Heckscher &  &  & Y & Auth. Ctrl. &  & Low \\
94 & Keiokaku Velodrome &  & Y & Y &  &  & Low \\
95 & Mena (given name) & Y &  &  &  &  & None \\
96 & Speed skating at the 1924 Winter Olympics – Men's all-round & Y &  &  &  &  & None \\
97 & List of Gintama. Shirogane no Tamashii-hen episodes & Y &  &  &  &  & None \\
98 & The Singing Hotel &  & Y &  &  &  & Low \\
99 & Kedrick & Y &  &  &  &  & None \\
100 & Arbinda Department &  & Y &  &  &  & Low \\\\
\caption{Complete listing of 100 random articles that were sampled from English Wikipedia and their Wikidata transclusion categories. Abbreviations: ``S.~only'' is ``Sitelinks only'' which means no transclusion; ``T. only'' is ``Tracking only'' which means there was transclusion  but it only generated tracking categories and no content; ``No shortdesc'' means that the article uses Wikidata's description in the mobile apps; ``Ib'' stands for ``Infobox''; ``Commons C.'' indicates ``Commons category'' template, which inserts a link to a Commons category in the page. Note that Infobox Anatomy is listed as metadata because it inserts metadata links as opposed to facts, though it is in the context of an infobox. Pages were evaluated in their current form on 31 July 2020 and they (or the templates they transclude) may have changed since then.}\label{tab:articlesample}
\end{longtable}
\end{scriptsize}

\end{document}